\documentclass{article}
\usepackage{hiph-preprint}
\usepackage{graphicx}
\usepackage{amssymb}
\usepackage{psfig}
\usepackage{epsfig}
\volnumber{22} \issuenumber{1} \edyear{2005}               
\frompage{000} \topage{000}                       
\recrevdate{1 January 2005}                       
\def\rightmark{Large Observed \vtwo as a Signature for Deconfinement}
\def\leftmark{W. A. Horowitz}
 
\newcommand{\be}{\begin{equation}}
\newcommand{\ee}{\end{equation}}
\newcommand{\bfig}{\begin{figure}}
\newcommand{\efig}{\end{figure}}
\newcommand{\bea}{\begin{eqnarray}}
\newcommand{\eea}{\end{eqnarray}}

\newcommand{\intd}{\mathrm{d}}

\newcommand{\raa}{$R_{AA}$ }

\newcommand{\raaphi}{$R_{AA}(\phi)$ }

\newcommand{\raaphiptcomma}{$R_{AA}(\phi;\,\eqnpt)$}
\newcommand{\vtwo}{$v_2$ }
\newcommand{\vtwocomma}{$v_2$}
\newcommand{\eqnraa}{R_{AA}}

\newcommand{\eqnvtwo}{v_2}

\newcommand{\auau}{$Au+Au$ }
\newcommand{\auaucomma}{$Au+Au$}

\newcommand{\cucu}{$Cu+Cu$ }

\newcommand{\rhopartcomma}{$\rho_{\textrm{\footnotesize{part}}}$}
\newcommand{\eqnrhopart}{\rho_{\textrm{\footnotesize{part}}}}

\newcommand{\eqnnpart}{N_{\textrm{\footnotesize{part}}}}

\newcommand{\eqntaa}{T_{AA}}

\newcommand{\eqnrhocoll}{\rho_{\textrm{\footnotesize{coll}}}}

\newcommand{\eqnncolltiny}{N_{\textrm{\tiny{coll}}}}

\newcommand{\eqndnslashdyabs}{dN_g^{abs}/dy}
\newcommand{\eqndnslashdyrad}{dN_g^{rad}/dy}

\newcommand{\dnslashdyabscomma}{$dN_g^{abs}/dy$}

\newcommand{\dnslashdy}{$dN_g/dy$ }
\newcommand{\dnslashdycomma}{$dN_g/dy$}
\newcommand{\eqndnslashdy}{dN_g/dy}
\newcommand{\as}{\alpha_s}
\newcommand{\alphas}{$\as$ }

\newcommand{\eqnalphas}{\as}

\newcommand{\pt}{$p_\perp$ }

\newcommand{\eqnpt}{p_\perp}

\newcommand{\highpt}{high-\pt}

\newcommand{\raapt}{$\eqnraa(\eqnpt)$ }
\newcommand{\raaptcomma}{$\eqnraa(\eqnpt)$}

\newcommand{\eqnrperpHS}{r_{\perp,HS}}
\newcommand{\eqnrperpWS}{r_{\perp,WS}}

\newcommand{\pizero}{$\pi^0$ }

\newcommand{\infinity}{\infty}

\newcommand{\rhopartvec}{\vec{x}_0+\hat{n}l}

\newcommand{\fig}[1]{Fig.~\ref{#1}}

\title{Large Observed \vtwo as a Signature for Deconfinement} 
\authors{ 
{W. A. Horowitz$^1$ %
\index{Horowitz, W. A.} 
}\\[2.812mm]
{\normalsize
\hspace*{-8pt}$^1$ Columbia University,\\ 
538 West 120$^{th}$ Street, New York, NY, USA\\[0.2ex]
}}
 
\abstract{We present a new plot for representing \raaphi data that emphasizes the strong correlation between \highpt suppression and its elliptic anisotropy. We demonstrate that existing models cannot reproduce the centrality dependence of this correlation. Modification of a geometric energy loss model to include thermal absorption and stimulated emission can match the trend of the data, but requires \dnslashdy values inconsistent with the observed multiplicity. By including a small, outward-normal directed surface impulse opposing energy loss, $\Delta\eqnpt\mathbf{\hat{n}}$, one can account for the centrality dependence of the observed \auau elliptic quench pattern. We also present predictions for \cucu reactions.}
\keyword{Heavy Ion Collisions, Jet Quenching, Deconfinement, RHIC}

\PACS{12.38.Mh; 24.85.+p; 25.75.-q}
 
\makeindex
\begin{document}
 
\maketitle
\section{Introduction}A theoretical model for RHIC mid- to high-\pt \raaphi should reproduce both the normalization as well as the azimuthal anisotropy of experimental results; its trend must follow the data on a \vtwo vs. \raa diagram. This is actually quite difficult due to the anticorrelated nature of \raa and \vtwocomma; previous models either oversuppressed \raa or underpredicted $v_2$ \cite{HGC,DFJ,shuryak}. In \fig{modelfailure} (a) and (b), we combine STAR charged hadron $\eqnraa(\eqnpt)$ and $\eqnvtwo(\eqnpt)$, PHENIX charged hadron $\eqnraa(\eqnpt)$ and \vtwo centrality, and PHENIX \pizero ($\eqnpt>4$ GeV) \raaphi centrality data \cite{STARraa}. We naively averaged the STAR and PHENIX $\eqnraa(\eqnpt)$ results to approximately match the \pt bins of their corresponding \vtwo measurements. We report the \raa and \vtwo modes of the PHENIX \pizero \raaphi data. The error bars provided are schematic only. 

Hydrodynamics cannot be applied to mid- to high-\pt particles due to the lack of equilibrium. Moreover, a naive application would highly oversuppress \raa due to the Boltzmann factors. Parton transport theory attempts to extend hydrodynamics' range of applicability to higher transverse momenta. The M\'olnar parton cascade (MPC) succeeded in describing the low- and intermediate-\pt \vtwo results of RHIC by taking the parton elastic cross sections to be extreme, $\sigma_t\sim45$ mb \cite{MPC}. One sees in \fig{modelfailure} (a) that for the MPC, in this instance run at approximately 30\% centrality, no single value of the controlling free parameter, the opacity, $\chi=\int dz\sigma_t\rho_g$, simultaneously matches the experimental \raa and \vtwocomma.

pQCD becomes valid for moderate and higher \pt partons, and models based on pQCD calculations of radiative energy loss have had success in reproducing the experimental \raapt data \cite{vitev}. These models use a single, representative pathlength; as such, they give $\eqnvtwo\equiv0$. To investigate the \vtwo generated by including pathlength fluctuations, we use a purely geometric (neglecting gluon number fluctuations) radiative energy loss model (GREL) based on the first order in opacity (FOO) radiative energy loss equation \cite{GLV2}; it has been shown that including the second and third order in opacity terms has little effect on the total energy loss \cite{multigluon}. The asymptotic approximation of this equation is $\Delta E^{(1)}_{rad}/E \propto (\eqndnslashdy) L^2$ \cite{wang2}. We thus use an energy loss scheme similar to \cite{DFJ}: $\epsilon = \Delta E_{rad}/E = \kappa I$. $\kappa$ is a free parameter encapsulating the $E$ dependence, etc.~of the FOO expression and the proportionality constant between \dnslashdy and \rhopartcomma. $I$ represents the integral through the 1D Bjorken expanding medium, taken to be $I=\int_0^\infinity\!\!\intd l\, l \frac{l_0}{l+l_0}\eqnrhopart(\rhopartvec)$, where $l_0=.2$ fm is the formation time. We consider only 1D expansion here because \cite{GVW} showed that including the transverse expansion of the medium has a negligible effect.

The power law spectrum for partonic production allows the use of the momentum Jacobian ($\eqnpt^f=(1-\epsilon)\eqnpt^i$) as the survival probability of hard partons. We distribute partons in the overlap region according to $\eqnrhocoll=\eqntaa$ and isotropically in azimuth; hence $\eqnraa(\phi;\,b) = \frac{\int\intd x\intd y\;\,\eqntaa(x,\,y;\,b)\left(1-\epsilon(x,\,y,\,\phi;\,b)\right)^n}{\eqnncolltiny}$, 
where $4\lesssim n\lesssim5$. The difference from using $n=4$ as opposed to $n=5$ is less than 10\%, and in this paper we will always use the former value. We evaluate \raaphi at 24 values of $\phi$ from 0-$2 \pi$ and then find the Fourier modes \raa and \vtwo of this distribution. Another method for finding \vtwocomma, not used here, assumes the final parton distribution is given exactly by \raa and \vtwocomma, and then determines \vtwo from the ratio $\eqnraa(0)/\eqnraa(\pi/2)$; this systematically enhances \vtwocomma, especially at large centralities. A hard sphere geometry is used for all our models, with $R_{HS}=6.78$ fm ensuring $<\eqnrperpWS^2>=<\eqnrperpHS^2>$. \fig{modelfailure} (a) shows that even with the HS-geometry-enhanced \vtwocomma, the GREL cannot recreate both \raa and \vtwo with a single parameter value. 
\section{Exclusion of Detailed Balance and Success of the Punch} 
\begin{figure}[htb]
\vspace{-.2in}
\epsfig{file=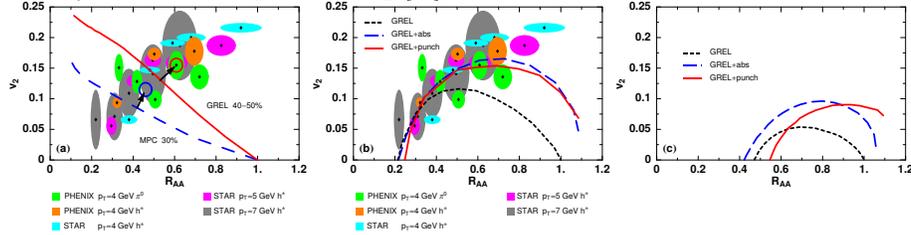,width=\columnwidth,angle=0}
\vspace{-.35in}
\caption{\footnotesize{(a) STAR $h^\pm$ data for 0-5\%, 10-20\%, 20-30\%,\ldots, and 40-60\%, PHENIX $h^\pm$ data for 0-20\%, 20-40\%, and 40-60\%, and PHENIX \pizero data for 10-20\%, 20-30\%,\ldots, 50-60\% centralities. Inability of previous models to fit the data. (b) Addition of thermal absorption or momentum punch to GREL; both fit the data, but absorption requires entropy-violatingly large \dnslashdyabscomma. (c) \cucu predictions for the three models.}}
\label{modelfailure}
\end{figure}

In \cite{wang2}, Wang and Wang derived the first order in opacity formula for stimulated emission and thermal absorption associated with the multiple scattering of a propagating parton, and found $\Delta E_{abs}^{(1)}/E \propto (\eqndnslashdy) L$. To model this we use $\epsilon = \frac{\Delta E_{rad}}{E}-\frac{\Delta E_{abs}}{E} = \kappa I-k I_2$, where $\kappa I$ is the same as in the GREL model, k is a free parameter encapsulating the proportionality constants in the absorption formula, and $I_2$ represents an integral through the 1D expanding medium: $I_2 = \int_0^\infinity\!\!\intd l\, \frac{l_0}{l+l_0}\eqnrhopart(\rhopartvec)$. $I_2$ has one less power of $l$ in the integrand; this permits a unique determination of the two free parameters, $\kappa=.5$ and $k=.25$ fits the 20-30\% centrality PHENIX \pizero \raaphi data point, and allows the model to duplicate the data as seen in \fig{modelfailure} (b). Taking the $\Delta E/E$ equations seriously, we invert them and solve for \dnslashdycomma; thus $dN^{rad}_g/dy \sim \kappa \frac{4E}{9\pi C_R \eqnalphas^3\tilde{v}_1} \frac{l_0 L}{l_0+L} \eqnnpart$, and $dN^{abs}_g/dy \sim k \frac{4E^2}{3\pi C_R \eqnalphas^2\tilde{v}_2} \frac{l_0 L}{l_0+L} \eqnnpart$, where $\tilde{v}_1$ and $\tilde{v}_2$ correspond to the bracketed terms in the energy loss and energy gain approximations of \cite{wang2}. For our fitted values of $\kappa$ and $k$, the choice of $E=6$ GeV, $L=5$ fm, and $\eqnalphas=.4$ gives $\eqndnslashdyrad\sim1000$ and $\eqndnslashdyabs\sim3000$ for most central collisions. For $E=10$ GeV, $\eqndnslashdyrad\sim1000$ and $\eqndnslashdyabs\sim9000$. The huge increase of $dN_g^{abs}/dy$ to values too large to fit the RHIC entropy data reflects the $E^2$ dependence of the Detailed Balance absorption. It seems the only way to have a large enough energy gain while maintaining $\eqndnslashdyabs\sim1000$ is to increase \alphas above 1. Note that these calculations were performed using a hard sphere nuclear geometry profile, which naturally enhances the produced \vtwo \cite{DFJ}.
\vspace{-.15in}

Building on the success of radiative energy loss in reproducing \raaptcomma, and supposing that latent heat, the bag constant, the screening mass, or other deconfinement effects might provide a small ($\sim1$ GeV) momentum boost to partons in the direction normal to the surface of emission, we created a new model based on the GREL model that includes a momentum ``punch,'' $\Delta\eqnpt$. After propagating to the edge of the medium with GREL, the parton's final, ``punched-up'' momentum and angle of emission are recomputed, giving a new probability of escape. Fitting to a single $(\eqnraa,\eqnvtwo)$ point provides a unique specification of $\kappa$ and punch magnitude. The results are astounding: one sees from \fig{modelfailure} (b) that a tiny, .5 GeV, punch on a 10 GeV parton reproduces the data quite well over all centralities. Fitting the PHENIX 20-30\% \pizero data sets $\kappa=.18$ and the aforementioned $\Delta\eqnpt=.5$ GeV. The size of the representative parton's initial momentum is on the high side for the displayed RHIC data; however, the important quantity is the ratio $\Delta\eqnpt/E$. Moreover, although the geometry used naturally enhances the \vtwocomma, we feel confident that when this model is implemented for a Woods-Saxon geometry, the necessarily larger final punch magnitude will still be relatively small. We expect the magnitude of this deconfinement-caused momentum boost to be independent of the parton's momentum; hence $\eqnvtwo(\eqnpt)$ will decrease like $1/\eqnpt$. Moreover, since $\epsilon$ is larger out of plane than in, a fixed $\Delta\eqnpt$ enhances $\eqnraa(\pi/2)$ more than $\eqnraa(0)$. These are precisely the preliminary trends shown by PHENIX at QM2005. Keeping the same values for $\kappa$, $k$, $\Delta\eqnpt$, etc.~as for \auaucomma, we show in \fig{modelfailure} (c) the centrality-binned \raa and \vtwo results for \cucu in the three geometric energy loss models. 
\section{Conclusions}
By failing to simultaneously match the \raa and \vtwo values seen at RHIC we discounted the MPC and pure GREL models. We showed that while including medium-induced absorption reproduces the \raaphi phenomena, it does so at the expense of inconsistent and huge \dnslashdycomma. But the addition of a mere 5\% punch created a RHIC-following trend. This impulse is small enough to be caused by deconfinement effects and future calculations should follow the \pt dependence of \raaphiptcomma. 
\section*{Acknowledgments}
The author wishes to thank B. Cole, M. Gyulassy, D. M\'olnar, and I. Vitev for their valued discussions. 

\vfill\eject
\end{document}